\documentclass{aastex62}

\usepackage{color}

\received{June 1, 2018}
\revised{January 7, 2018}
\accepted{\today}
\submitjournal{AJ}

\shorttitle{TD ANN}
\shortauthors{Peng et al.}

\begin{document}

\title{Optical Transient Object Classification in Wide Field Small Aperture Telescopes with Neural Networks}

\author[0000-0001-6623-0931]{Peng Jia}
\affil{College of Physics and Optoelectronics,  Taiyuan University of Technology, Taiyuan, 030024, China}
\affiliation{Department of Physics, Durham University, South Road, Durham DH1 3LE, UK}
\affiliation{Key Laboratory of Advanced Transducers and Intelligent Control Systems, Ministry of Education and Shanxi Province, Taiyuan University of Technology, Taiyuan, 030024, China}
\email{robinmartin20@gmail.com}

\author{Yifei Zhao}
\affiliation{College of Physics and Optoelectronics,  Taiyuan University of Technology, Taiyuan, 030024, China}

\author{Gang Xue}
\affiliation{College of Physics and Optoelectronics,  Taiyuan University of Technology, Taiyuan, 030024, China}

\author{Dongmei Cai}
\affiliation{College of Physics and Optoelectronics,  Taiyuan University of Technology, Taiyuan, 030024, China}



\begin{abstract}
Wide field small aperture telescopes are working horses for fast sky surveying. Transient discovery is one of their main tasks. Classification of candidate transient images between real sources and artifacts with high accuracy is an important step for transient discovery. In this paper, we propose two transient classification methods based on neural networks. The first method uses the convolutional neural network without pooling layers to classify transient images with low sampling rate. The second method assumes transient images as one dimensional signals and is based on recurrent neural networks with long short term memory and leaky ReLu activation function in each detection layer. Testing with real observation data, we find that although these two methods can both achieve more than $94\%$ classification accuracy, they have different classification properties for different targets. Based on this result, we propose to use the ensemble learning method to further increase the classification accuracy to more than $97\%$.\\
\end{abstract}

\keywords{techniques: image processing --- methods: numerical  --- surveys}


\section{Introduction} \label{sec:intro}
A synoptic survey carried out by several telescopes with small apertures and wide field of view (WFSAT) is a cost-effective way for time domain astronomy observations. Several WFSATs equipped with automated instruments can be placed in several different sites around the world, or in the same site pointing to different directions to fulfill a variety of astronomical survey tasks, including discovery of optical counterpart of gamma ray bursts, gravitational-wave sources and asteroids, photometry of extra solar planets, variable stars and astrometry of near earth objects \citep{Kaiser et al.(2002), Burd et al.(2005), Pollacco et al.(2006), Schildknecht(2007), Yuan et al.(2008), Drake et al.(2009), Tonry(2011),Law et al.(2012), Yuan et al.(2014),Sun et al.(2014)}.\\
For surveys carried out by WFSATs, necessary trade off should be made between the field of view, the magnitude limitation and the cadence interval for different scientific aims as discussed by \cite{Tonry(2011)}. Generally speaking, limited by contemporary technology of CCD camera, images obtained by WFSATs usually have relative large pixel scale (several arcsec per pixel). Considering the image quality of optical system and the size of seeing disc (normally around 1 to 2 arcsec), these images are often critically sampled or even under-sampled. Although under-sampling is not a critical problem to the system's sensitivity, it will bring problems to our transient detection task.\\
In this paper, we consider WFSATs which are working in white light mode (no filters) to maximize their detection ability. For images obtained by these telescopes, we are interested in newly discovered transients and moving transients instead of transients with variable magnitude.  To fulfill our detection request, we directly scan original images and compare the scanned results with the star catalogue of \citet{Hog et al.(2000)}, to find transients instead of scanning difference images \citep{Zackay et al.(2016)}. With this observation mode, the transient detection here refers to extracting images of candidate astronomical transients from observational images and then classifying these candidates between real sources and artifacts. The first step is normally carried out by SExtractor \citep{Bertin(1996)} or other source extracting algorithms. Because the under-sampling problem will reduce pixel number of candidate transient images, real transients with small pixel number are very likely to be confused with artifacts. Artifacts are images generated by the statistical fluctuations of the background, cosmic rays and bad CCD pixels, which will also be extracted from the observation image by source extracting algorithms. A lot of false alarms will be triggered by these artifacts, if we directly use source extracting results as transient sources. These false alarms will waste time of follow up telescopes and reduce overall observation efficiency.\\
To increase detection accuracy, candidate transient images will be cut from the observation image as stamp images and then machine learning algorithms will be used to classify transients from artifacts. Random forest (RF) with features designed carefully for each telescope can give promising results and becomes the major method for general purpose surveying projects \citep{Brink et al.(2013), Goldstein et al.(2015), Cao et al.(2016), Masci et al.(2017), Lin et al.(2018)}.\\
For observations carried out by WFSATs, because some features are instrumentation-related, we need to set up these features carefully for every WFSAT. Besides, when the observation condition changes, we need to review the new observation data and adjust parameters of these features for every WFSAT to keep the classification algorithm effective. It will cost a lot of engineering time. Because transient detection is a temporal intensive task (most of them need to be followed up by other survey telescopes or observed through spectroscopy or precision photometry and astrometry), RF methods will place a very intensive workload level for pipeline maintenance and will slow down the transient detection efficiency.\\
In recent years, learning features from data by artificial neural network (ANN) \citep{Lecun et al.(2015), He et al.(2015), Goodfellow et al.(2016)} has become a candidate method for transient detection \citep{Theodoridis(2001), Djorgovski et al.(2011), Graff et al.(2014), du Buisson et al.(2015)}. For transient detection in WFSATs, ANN can learn complex features directly from the observation data. When the observation condition changes, we could use new data labeled by human experts to train the ANN for each WFSAT effectively through transfer learning \citep{Sinno(2009)} or increment learning \citep{Xiao et al.(2017)}. Because ANN has advantages mentioned above, in this paper we propose two ANN based classification methods. The first method is a convolutional neural network (CNN) based method \citep{Mahabal et al.(2008), Krizhevsky et al.(2012),Cabrera-Vives et al.(2017), Wright et al.(2017)} which does not include any pooling layers, making it suitable to classify low sampling images. The second method  assumes the images of transients as one dimensional signal and uses a recurrent neural network with long short term memory (RNN) for transient classification \citep{Sak et al.(2014)}. We will analyze different properties of transient and artifact images captured by WFSATs in Section \ref{sec:data}. In Section \ref{sec:dnn}, we will introduce our methods and compare their classification performance in Section  \ref{sec:results}. We will make our conclusions and anticipate our future work in Section \ref{sec:con}.\\ 

\section{Data properties of transient object images captured by WFSATs} \label{sec:data}
\subsection{The PSF of WFSATs and Its Impact to Classification Algorithm Chosen}\label{subsec:PSF}
The imaging process in a telescope can be modeled by the following equations:
\begin{equation} \label{eq:1}
I(x,y,t_0)=\int_{t_0}^{t_0+\delta _t}\left[O(x,y,\tau)\ast PSF(x,y,\tau)\right]_{pix}+N(x,y,\tau)d\tau,
\end{equation}
where $I(x,y,t_0)$ is the image obtained in $t_0$ with exposure time of $\tau$, $O(x,y,\tau)$ is the `original image', $PSF(x,y,\tau)$ is the PSF of the whole observation system, $N(x,y,\tau)$ is the noise which includes the background noise and read-out noise, $\ast$ is convolution operator and $\left[\right]_{pix}$ is the pixel response of CCD.\\
WFSATs are optimized to search sky with high cadence for transient discovery. This scientific aim requires WFSATs to have a wide field of view (bigger than 1 square degree), which will always lead to telescopes with small aperture (around or less than 1 meter) and refract components to correct aberrations and atmospheric dispersion. The optical design of WFSATs make them obtain images with good quality in a cost effective way (the optical aberrations can be greatly reduced and the atmospheric turbulence becomes the major contribution to the image quality). However, contemporary technology can not provide science cameras with low noise, high speed and very large number of pixels at the same time, leading to low spatial sampling rate in the image plane of WFSATs.\\ 
Considering good image quality and low sampling rate in the image plane, the model of imaging process in WFSATs can be expressed as:\\
\begin{equation} \label{eq:2}
I(x,y,t_0)=\left[O(x,y,t_0)\ast PSF_{Atm}(x,y,t_0)\right]_{pix}+N(x,y,t_0),
\end{equation}
$PSF_{Atm}(x,y,t_0)$ is the point spread function of atmospheric turbulence during exposure time $t_0$, others are integral results of the parameters defined in equation \ref{eq:1}. For ordinary sites, the atmospheric turbulence will introduce a PSF with full width half magnitude (FWHM) of around 1 to 2 arcsec \citep{Roddier(1981)} and can be less than 1 arcsec in good observation sites \citep{Jee(2011)}. Because the pixel scale is around several arcsecs in images captured by WFSATs, the FWHM of dim transient images will be around or smaller than 3 pixels.\\
However for bright targets captured by WFSATs, we find that the size of their images is not fixed by the FWHM of the atmospheric turbulence induced PSF and can be more than several arcmins. This is caused by the aureole region of the PSF \citep{Racine(1996)}, which comes from scatter of the atmosphere and the telescope \citep{Sandin(2014)}. The energy of star in the aureole region is too small to be assumed as part of PSF for ordinary targets, but it will still change the statistical property of the gray values in this region. For data obtained by a particular WFSAT, we can design a machine learning algorithm to exploit the entire information contained within the whole candidate transient image (up to tens of arcsec) for transient classification, because we assume that the unknown relation between the gray scale values in the core region and in the aureole is stable.\\
\subsection{Data Properties and Transient Detection Aims}
\begin{figure*}
\includegraphics[width=0.3\textwidth]{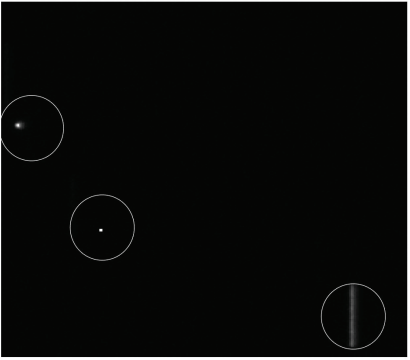}
\caption{Part of the observation image. We place a point-like artifact in the middle of the original image. This figure shows a point-like transient (in the left circle), an artifact caused by a hot pixel (in the middle circle) and a streak-like transient (in the right circle). The gray scale of this figure is rescaled by the zscale algorithm in ds9 \citep{Joye and Mandel(2003)}.}\label{fig1}
\end{figure*}

\begin{figure*}
\includegraphics[width=0.9\textwidth]{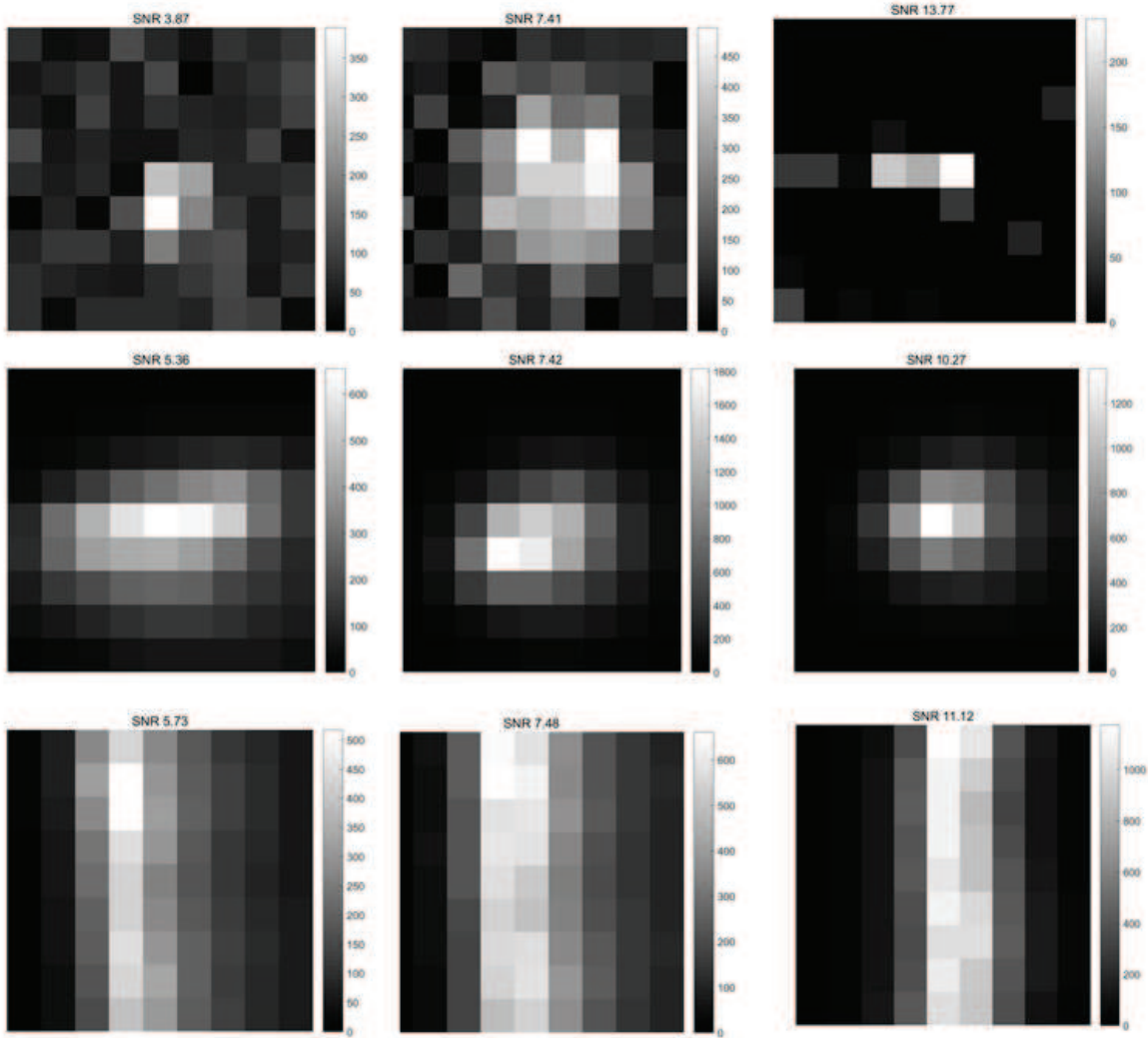}
\caption{Examples of stamp images obtained from observation data. Images in the first row are artifacts, images in the second row are point-like transients and images in the third row are streak-like transients with different SNR. All these targets have been cross-matched with the star catalogue and checked by human experts. The gray scale of these images have been rescaled by the zscale algorithm in ds9 \citep{Joye and Mandel(2003)} and the color bar shows the ADU (Analog-Digital Units) of these images.\label{fig2}}
\end{figure*}

In this paper, we use Xuyi Schmidt Telescope \footnote{This telescope is operated by the Near Earth Objects Research Group of Purple Mountain Observatory, China Academy of Science and the information of this telescope can be found in \url{http://english.pmo.cas.cn/rs/fs/200909/t20090918_39134.html}} to obtain the observation data. This telescope is a classical Schmidt telescope with equatorial fork mount. It has an aperture of 1200 mm and a Schmidt corrector of 1040 mm. The image quality of this telescope is $80\%$ energy inside $0.57\arcsec$ and the seeing disc is less than $1 \arcsec$. The observation data used in this paper has a pixel scale of $1.705 \arcsec$. It is a classical WFSAT according to its optical design, aperture size, field of view, pixel scale and the FWHM of the atmospheric turbulence induced PSF. We have compared images of different targets and decided to extract images with $9\times9$ pixels (around $15 \arcsec \times 15 \arcsec$) as transient candidate images for classification.\\
The observation data we used in this paper are dedicated for geostationary earth orbit (GEO) space debris survey, which has the same observation strategy as those used for comets and asteroids detection \citep{Sun2 Zhao(2014), Sun et al.(2015), Waszczak(2017), Zhai(2018)}. With this observation strategy, images of GEO objects will be point-like and that of celestial objects and other transients will be streak-like. We test our algorithm with this data set as proxy in this paper and our future surveys will be working in sidereal-tracking mode. In the sidereal-tracking mode, the telescope will be used for early discovery of transient sources and moving targets at the same time \citep{Molotov et al.(2008)}, which means we need to design an algorithm to classify candidate transient objects between artifacts, point-like transients and streak-like transients. As shown in Figure \ref{fig1}, there are morphological differences between point-like and streak-like transient candidates. For transient images and artifacts, as we discussed in Section \ref{subsec:PSF}, because transient images are modified by the PSF while artifacts are not, classification based on the unknown relation between these images are another task for our algorithm.\\
We scan whole frame of images with SExtractor \citep{Bertin(1996)} to detect transient candidates. The minimal pixel number of connected area and detection threshold are very small (2 pixels and 1.1 to background) to make sure that all transients can be extracted. According to extraction results, we cut stamp images with size of $9\times9$ pixels from observation images. Normally each stamp image will contain one transient candidate and the brightest pixel is around its center. We then cross-match all the streak-like images with the star catalogue of \citet{Hog et al.(2000)}, calculate the orbit of the point-like images with several frames and check their orbit with the space debris or satellite orbit data. At last we set the remaining stamp images as artifacts. The total number of stamp images is 9000 for streak-like transients (stars), 90 for point-like transients (GEO space debris or satellites) and 1000 for artifacts. The images we observed have short exposure time (5 sec per frame) which causes these images have relatively low SNR. We define SNR of a stamp image according to \citet{Staley(2014)}:\\
\begin{equation} \label{eq:3}
SNR=\frac{I-N\times \bar B}{\sqrt{N\times \sigma_B}},
\end{equation}
where $I$ is the total flux of the star, $\bar B$ is the mean value of the background, $\sigma_B$ is the variance of the background and $N$ is the total pixel number of the stamp image. We have calculated the SNR of all the stamp images. The statistical results are shown in Table \ref{tab3}. We can find that they have relatively small SNR and low sampling rate as shown in figure \ref{fig2}.\\ 
\startlongtable
\begin{deluxetable}{cccc}
\tablecaption{The SNR of different targets. $SNR_{mean}$, $SNR_{min}$ and $SNR_{max}$ are the mean value, minimal and maximal values of SNR for different targets. \label{tab3}}
\tablehead{
\colhead{Targets} & \colhead{$SNR_{mean}$} & \colhead{$SNR_{min}$}& \colhead{$SNR_{max}$}
}
\startdata
Point-like & 7.47 & 5.36 & 12.16\\
Streak-like & 6.91&3.10 &11.12 \\
Artifacts & 4.29 & 3.01 & 47.15
\enddata
\end{deluxetable}
\section{ANN Based Transient Detection Methods} \label{sec:dnn}
\subsection{CNN for Transient Classification}
\begin{figure*}
\includegraphics[width=0.7\textwidth]{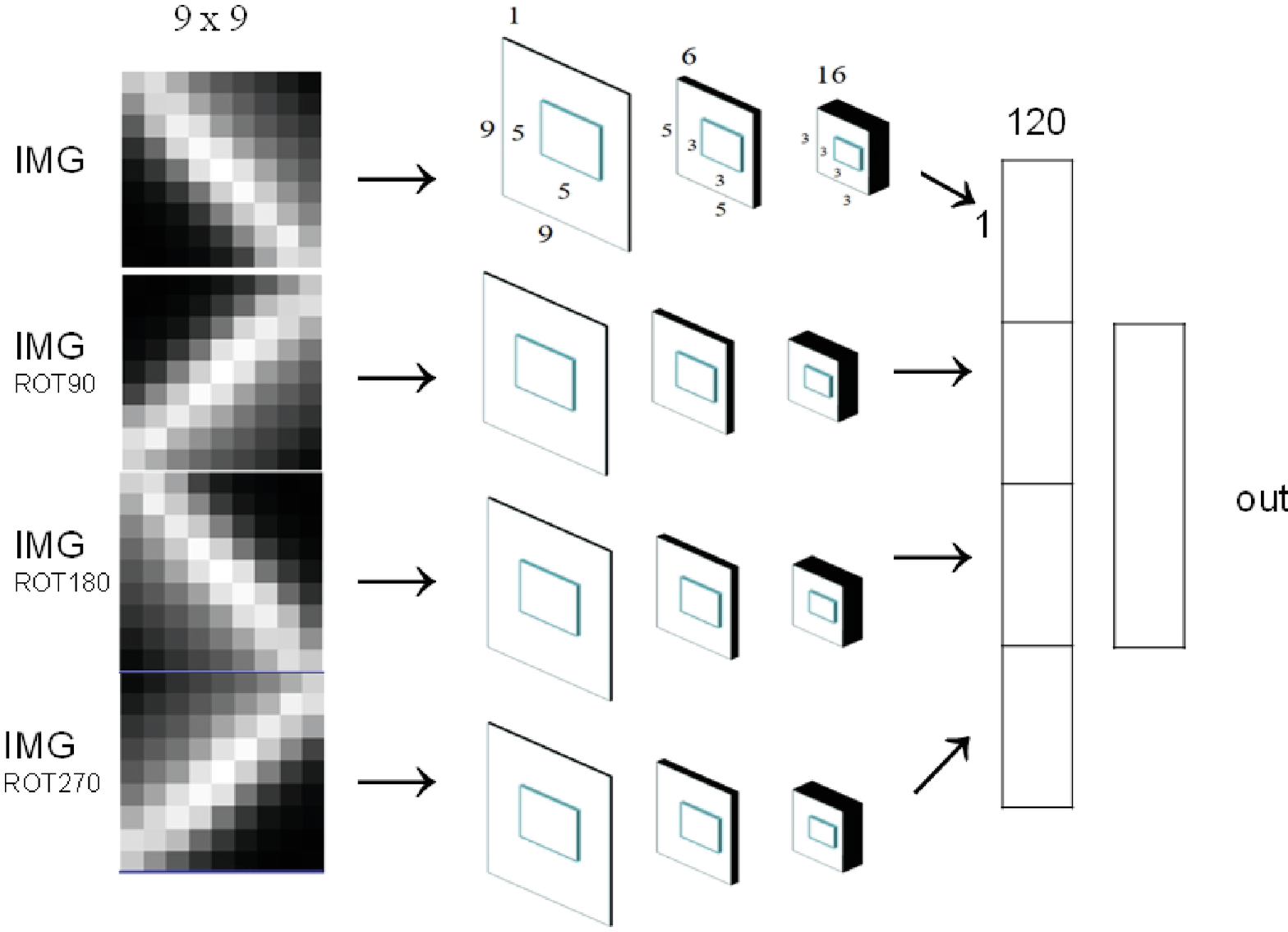}
\caption{The architecture of CNN used in this paper. IMG is the original image data and the original image will be sent into four parallel CNNs. Each CNN includes three levels of convolutional layer. The input size, convolutional kernel size, the stride size and the output size are shown in the up-most CNN.}\label{fig3}
\end{figure*}

Transient classification can be viewed as a classical image classification task. We use CNNs since they have proved powerful for image classification. This power arises from the ability of CNNs to learn translationally invariant features in trainable convolutional kernels. As shown in Figure \ref{fig3}, our network consists of three convolutional layers, the first of which has a 5x5 trainable convolutional kernel, whilst the last two have 3x3 convolutional kernels. The outputs of each of these convolutions is activated using Leaky ReLU with a negative gradient of 0.01. We use four identical sets of convolutional layers to learn each rotation of the input image separately. The outputs of each of these convolutional layers is flattened, concatenated and passed through two dense layers with 120 and 84 neurons respectively, each of which is activated using Leaky ReLU with a negative gradient of 0.01. We also add batch-normalization in the final fully connected layer. The output of the final dense layer is compressed to three classifications with softmax activation. This network has 86575 parameters and is highly influenced by Deep-HiTS  \citep{Cabrera-Vives et al.(2017)}, but was modified for the task in hand.\\
\subsection{RNN for Transient Classification}
\begin{figure}
\includegraphics[width=0.6\textwidth]{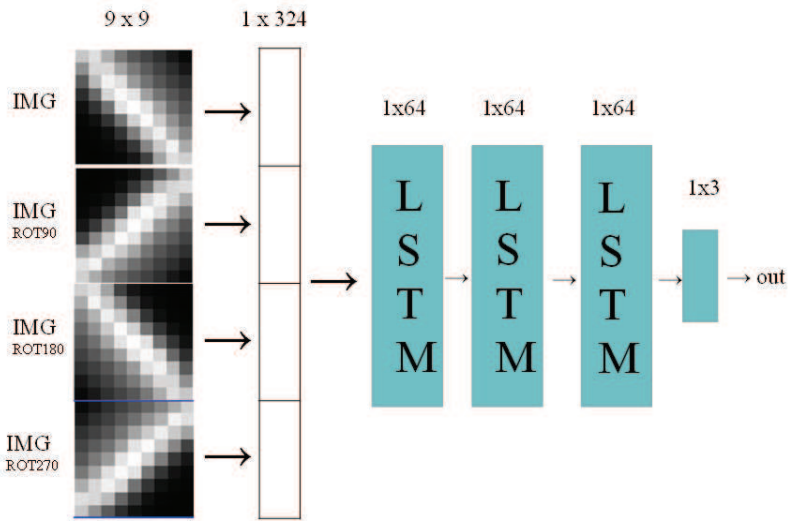}
\caption{The architecture of RNN based DNN in this paper. IMG is the orignal image data, LSTM is the LSTM layer and out is the output of the last fully connected layer and a softmax layer.}\label{fig4}
\end{figure}

RNN is another type of ANN \citep{Rumelhart et al.(1986)} and it has been used as a competitive alternative to CNN in image related tasks \citep{Graves et al.(2007), Visin et al.(2015)}. In this paper, we use RNN with long short term memory (LSTM) for the transient classification task. LSTM is a particular RNN architecture and it has an input gate, an output gate and a forget gate. The LSTM has several hidden dimensions to learn the long-term dependencies in the input sequences (gray value variations in the transient candidate image for our application).\\
As shown in Figure \ref{fig4}, the RNN used in this paper includes one input layer, three LSTM layers and one output layer. Candidate transient images with size of $9\times9$ pixels will be rotated with 3 orthogonal directions ($90^{\circ}$, $180^{\circ}$ and $270^{\circ}$) , stretched into a vector with dimension of $1\times81$, stitched together and sent into the input layer. The input layer is a fully connected layer with 324 inputs and 9 outputs which are fed into three consecutive LSTM layers. Each of the LSTM layer has 64 hidden neurons, 9 neurons as input and  9 neurons as output respectively. The Leaky-ReLU function with a negative gradient of 0.01 is used as activation function of the forget gate, the input gate and the output gate in all LSTM layers. The output layer compresses the output of the last LSTM layer into three dimensions and uses a softmax function as activation function to output the classification results. The total number of parameters is 85955 in this RNN. This RNN can hold a complex enough model for the transient detection task. Although the risk of over-fitting will increase, we will use drop out to reduce the risk of over-fitting as discussed in Section \ref{sec:results}. \footnote{The complete code of the CNN and the RNN used in this paper can be downloaded from \url{http://aojp.lamost.org}.}.\\
\section{Classification Results} \label{sec:results}
\subsection{Training Process}
\begin{figure*}
\includegraphics[width=0.8\textwidth]{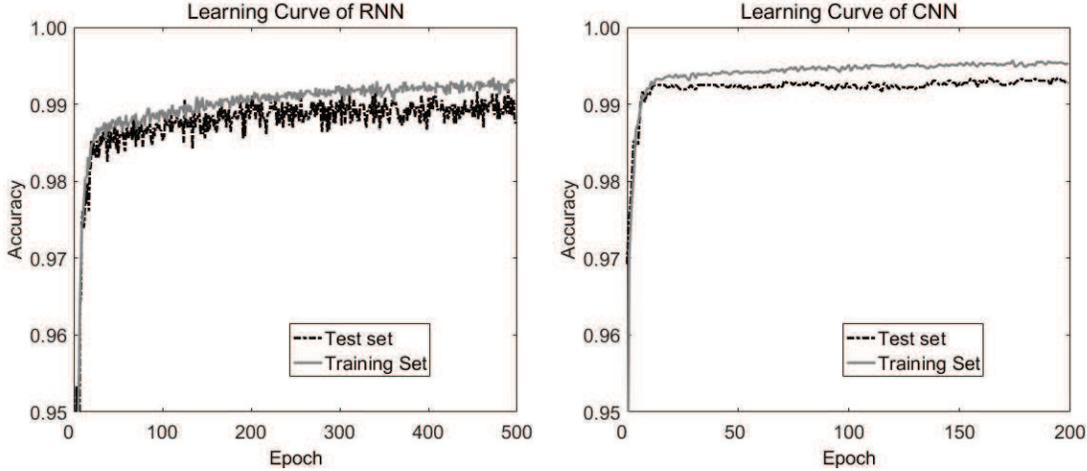}
\caption{The learning curve of the CNN and the RNN. The accuracy of the CNN will converge after 10 epochs. The accuracy of the RNN is still increasing after 200 epochs and will converge after 300 epochs.}\label{fig5}
\end{figure*}

Because the number of parameters is large in the RNN and the CNN, while the quantity of our stamp images is small and the number of stamp images with different labels is not balanced, we use the following tactics during training process:\\
1. When we read the stamp images into our training and testing dataset, we will randomly rotate these stamp images to generate several new stamp images of streak-like, point-like transients and artifacts with different rotation angles. The neural network will better generalize through this data augmentation step.\\
2. To prevent poor generalisation of the ANN, we use early-stopping strategy during training process.\\
3. According to our experience, we set both the RNN and the CNN with drop-out rate of 0.1 (set $10\%$ of neurons to be zero during training) to prevent over-fitting.\\
We build the CNN and the RNN with Pytorch and train them on a GTX 970 graphics processor unit (GPU). The training data and testing data are sent into the CNN or the RNN by epochs. All the transient stamp images will be mixed together and in each epoch, we randomly select 6000 streak-like transient stamp images, 6000 point-like transient stamp images and 6000 artifacts stamp images as training set and the same number of different images as test set. We divide each epoch into 180 batches and 100 images of different transients are mixed in each batch. Both the RNN and the CNN are optimized by the Adam algorithm \citep{Kingma(2014)}. The learning rate for the CNN is fixed as $10^{-4}$. For the RNN, we set the start learning rate as 0.01 to make it converge fast and decrease the learning rate after every 10 epochs (1800 iterations) with equation \ref{eq:12} to prevent oscillation around the optimal value,\\
\begin{equation} \label{eq:12}
L_{n}=0.5\times L_{n-1} \pmod{10},
\end{equation}
where $L_{n}$ is the learning rate in epoch $n$ and $L_{n-1} \pmod{10}$ stands for moduling calculation. Figure \ref{fig5} shows the accuracy of the CNN and the RNN as a function of iterations for the training set and the test set. To better show the performance of different classification methods, we train each neural network with 500 epochs. For these two neural networks, we find that the CNN will converge after 10 epochs (1800 iterations and cost around 27 seconds) and we use 10 epochs to train the CNN afterwards. For the RNN, the accuracy is still increasing after 200 epochs and will converge after 300 epochs. However, the increment of the accuracy is very small after 200 epochs. Besides, becasue we have only limited transient candidate images as training set, we use the early-stopping tactic to prevent poor generalisation of the RNN and use 200 epochs (36000 iterations and cost around 900 seconds with our hardware) to train the RNN.\\
\subsection{Comparison of classification results of the CNN and the RNN}\label{sec:comparison}
\subsubsection{6-fold cross-validation results}\label{subsec:6fold}
\begin{figure*}
\includegraphics[width=0.99\textwidth]{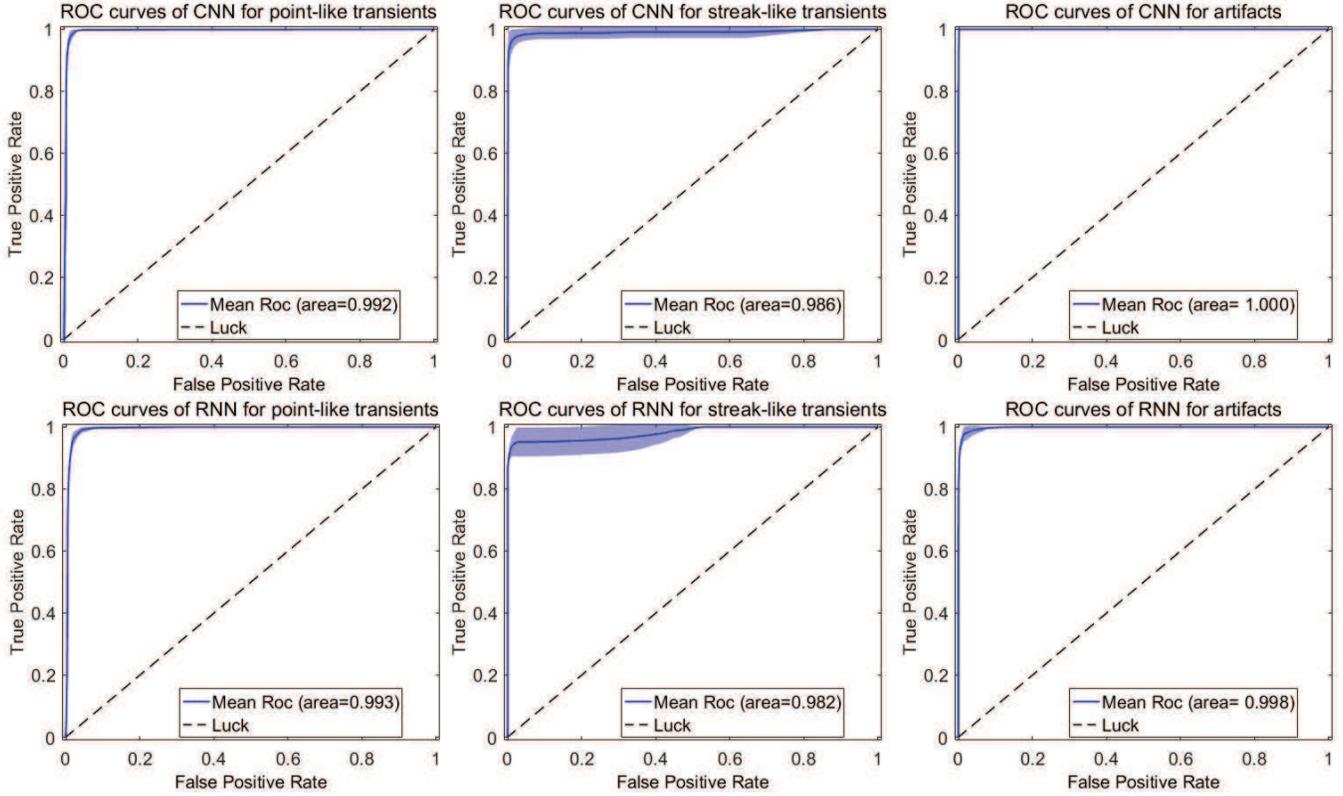}
\caption{The classification performance of the CNN and the RNN for point-like transients and streak-like transients in the validation set. This plot is the TPR and FPR relation, which is also called the receiver operation characteristic curve (ROC). The top three figures are ROCs for CNN and the bottom three figures are ROCs for RNN. The mean mean area under curve (AUC) of artifacts is 0.998 for RNN and 1.000 for CNN. The mean AUC of point-like transients is 0.993 for RNN and 0.992 for CNN. The mean AUC of streak-like transients is 0.982 for RNN and 0.986 for CNN. The filled areas show the distribution of the ROCs from all the validation sets.}\label{fig6}
\end{figure*}

\begin{figure*}
\includegraphics[width=0.99\textwidth]{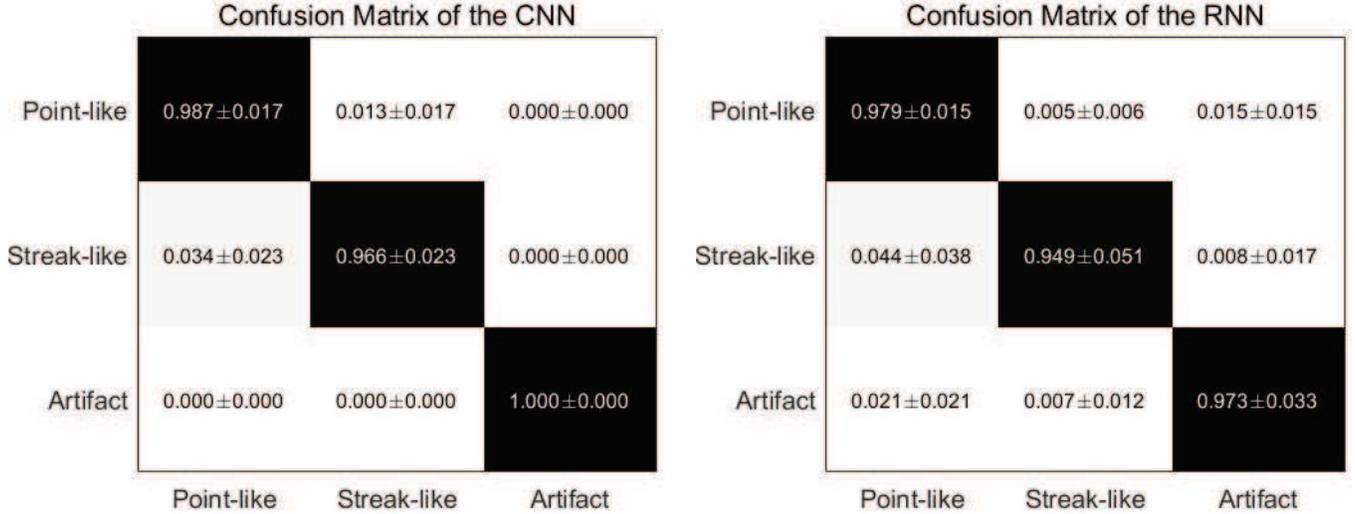}
\caption{The statistical confusion matrix of the classification results of the CNN and the RNN between different transient candidates in the 6-fold cross validation.}\label{fig7}
\end{figure*}

We test our classification algorithms with the observation data discussed in Section \ref{sec:data} by the 6-fold cross-validation method. We firstly split the whole data set into 6 disjoint data sets through random sampling. Then we initialize our CNN or RNN with random weights and train the CNN or RNN with five of the data sets and use the other one data set as the test set. Then we train the CNN or the RNN with initiation of random weights 5 times with 5 different data sets as training set and the other data set as test set. We use true positive rate (TPR) and false positive rate (FPR) to show the performance of our RNN as defined in equation \ref{eq:13},\\
\begin{eqnarray}\label{eq:13}
\begin{array}{cc}
TPR=\frac{TP}{TP+FN},\\
FPR=\frac{FP}{FP+TN},
\end{array}
\end{eqnarray}
where TP is true positives (number of correctly classified positives), TN is true negatives (number of correctly classified negatives), FP is false positives (number of incorrectly classified positives) and FN is false negative result (number of incorrectly classified negatives).\\
We will firstly use the receiver operation characteristic curve (ROC) and the area under the curve (AUC) to compare the performance of the CNN and the RNN. The ROC illustrates the classification ability of the algorithm and the AUC gives a quantified diagnosis standard. The AUC has a maximal value of 1 and minimal value of 0, and its value measures the discrimination ability of a particular classification algorithm. The bigger the AUC value, the better the classification algorithm is. Since the CNN and the RNN will classify stamp images into three different kinds, we will show the ROCs of the test set for different transients in Figure \ref{fig6}. As shown in this figure, the mean AUCs of the RNN for the 6-fold cross validation are 0.982 for streak-like transients, 0.993 for point-like transients and 0.998 for artifacts. While the mean AUCs of the CNN are 0.986 for streak-like transients, 0.992 for point-like transients and 1.000 for the artifacts.\\
To better understand the performance of the RNN and the CNN, we show the statistical confusion matrix in Figure \ref{fig7}. In the confusion matrix, we calculate the mean value and the variance from the results of the 6-fold validation. For the CNN, we can find that the main classification error lies in the misclassification between streak-like transients and point-like transients. For streak-like transients classification task, the CNN can achieve around $96.6\%\pm 2.3\%$ accuracy. For point-like transients classification task, the CNN can achieve around $98.7\%\pm 1.7\%$ accuracy. However, we can notice that the CNN can achieve almost $100 \%$ accuracy for artifacts classification. This is probably caused by the fact that the image of artifact is not modulated by the PSF and it has much smaller size than that of the transient, which makes it easier to be classified by the CNN. However, we should mention that the artifacts are introduced by many different processes and the training set used here may not include all artifacts. So in real applications, the accuracy of artifacts classification may vary. Besides we output the classification results directly according to the output of softmax layer and we can adjust the output threshold according to different completeness and reliability requirements.\\
For comparison, we also show the RNN classification results in Figure \ref{fig7}. We can find that the RNN is slightly better at classification between the point-like transient and the streak-like transient. However, the RNN will classify around $3\%$ transients into artifacts, which will become a problem for transient detection. This may be caused by the similarity between the point-like transient images and the artifact images. To check this problem, we will further compare the performance of these two algorithms with newly observed data as validation set.\\

\subsubsection{Classification results with newly observed data}
\label{newlydata}
\begin{figure*}
\includegraphics[width=0.99\textwidth]{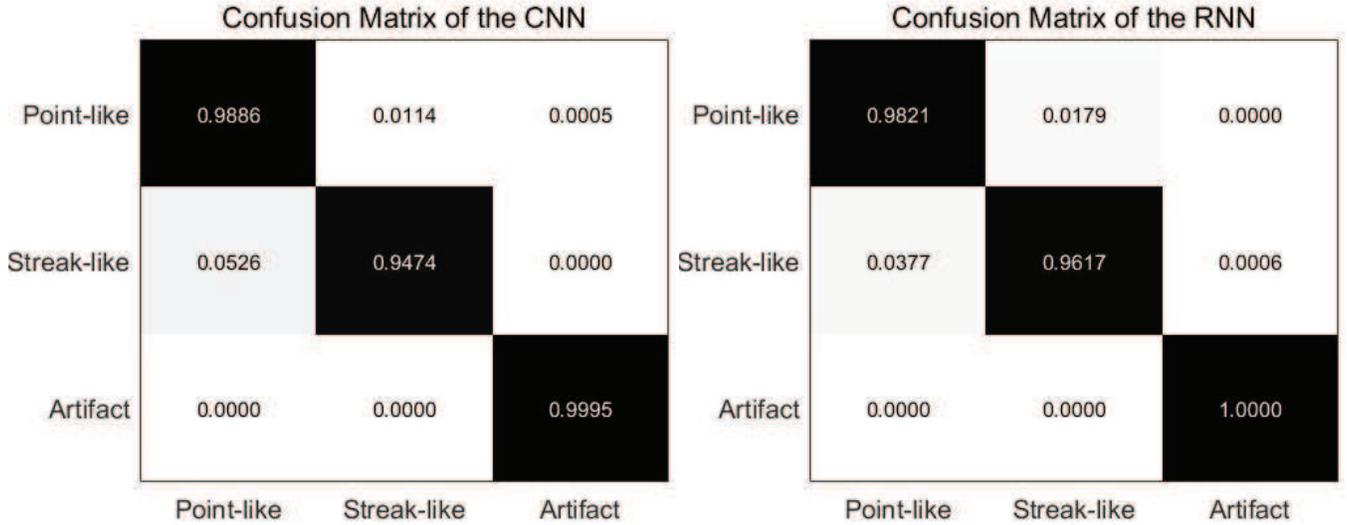}
\caption{The confusion matrix of the classification results of the CNN and the RNN between different transient candidates in the newly observed data. We use 4 decimals here to better show classification results.}\label{fig8}
\end{figure*}

We test our algorithms with stamp images from a newly observed data set. The observation mode is the same as that of the data set we used for neural network training. We obtain transient candidates with the same source extract method and check these transient candidates by human experts according to the star catalog \citep{Hog et al.(2000)}, the satellite and the space debris orbits. The mean SNR of images in the validation set are around 18.49 for streak-like transients, 9.12 for point-like transients and 4.29 for artifacts. We use all the labeled images as training set to retrain the CNN and the RNN and use the retrained ANNs to classify transient candidates in this validation set. The confusion matrix of the classification results in the validation set are shown in Figure \ref{fig8} and the ROC curves are shown in Figure \ref{fig9}. \\
First of all, the AUCs of this newly observed data are bigger than the mean AUCs of the validation set we discussed in Section \ref{subsec:6fold}. We have checked the 6-fold validation set and have found that there are several transient candidate images that are hard to be classified by both of these two ANNs. However, we will not remove these images from the 6-fold validation set to test the robustness of these two ANNs. Secondly in this newly observed data set, as shown in the confusion matrix, both of these ANNs work well for transient classification, however, we can find that the classification accuracy of the CNN drops down while that of the RNN increases, which indicates that the classification ability of these ANNs is different. These differences inspire us to design an algorithm to make better use of both of these ANNs.\\
\subsection{Ensemble classification method} \label{subsec:hybclas}
\begin{figure*}
\includegraphics[width=0.99\textwidth]{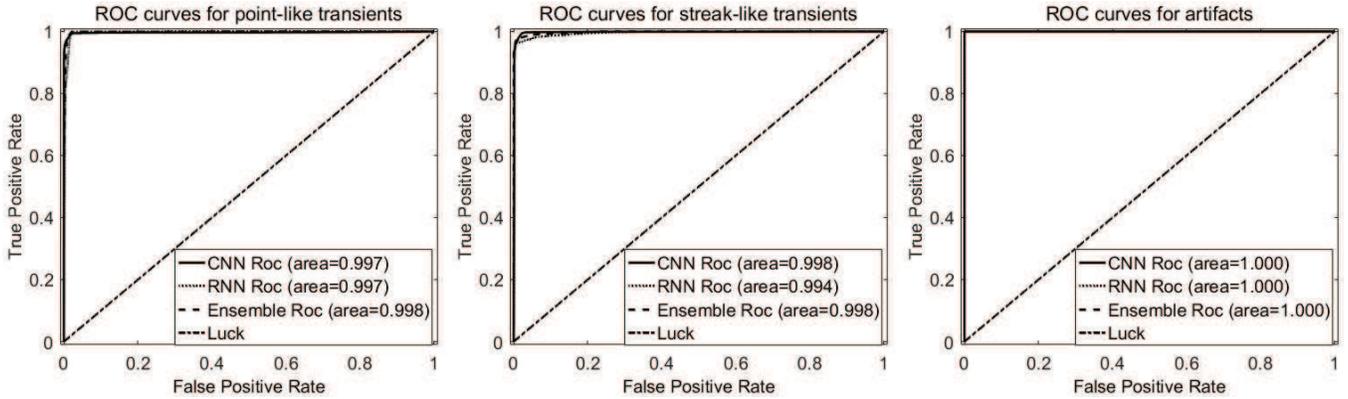}
\caption{The classification performance of the RNN, the CNN and the ensemble classification methods. The left figure shows the receiver operation characteristic curve of the artifacts, the middle figure shows the receiver operation characteristic curve of the streak-like transients and the right figure shows the ROC of the point-like transient. We can find that the AUC of the ensemble method is bigger than that of other methods, which shows that the ensemble method has better performance in real application.}\label{fig9}
\end{figure*}

\begin{figure*}
\includegraphics[width=0.5\textwidth]{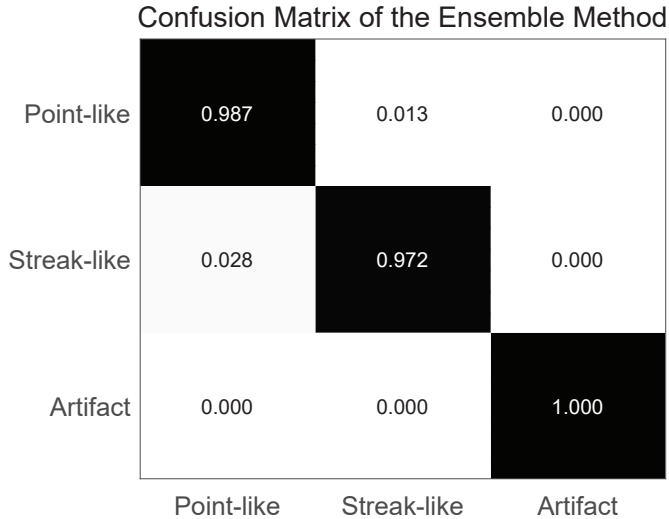}
\caption{The confusion matrix of the ensemble method of the CNN and the RNN between different transient candidates in the newly observed data. }\label{fig10}
\end{figure*}

Because the CNN and the RNN have different classification performance, we can build a ensemble learning algorithm to achieve more stable performance in real applications. There are several ensemble learning methods that can achieve this requirement, such as AdaBoost \citep{kegl2013} or bagging \citep{Breiman L.(1996)}. In this paper, we use the bagging method. We will load the trained CNN and RNN model and sum outputs from last linear layer of the CNN and the RNN. Then a softmax layer will be used to give the classification results. The ensemble method is used to classify the newly observed data discussed in subsection \ref{newlydata}. We can find that the classification accuracy is increased to more than $97{\%}$ as shown in Figure \ref{fig10}. The misclassification rate between different transients and artifacts is reduced, with the cost of a slightly improved misclassification between the point-like transient and the streak-like transient. The ROC curves of the CNN, RNN and ensemble methods are shown in figure \ref{fig9} and it shows that the ensemble method is more stable for transient classification. It indicates us to use ensemble learning in our future work.\\
\section{Conclusions and Future Work} \label{sec:con}
Fast and accurate classification of transient candidates into different classes is important for transient discovery tasks carried out by WFSATs. In this paper, we propose two new ANN based methods for transient image classification. Both of these methods can classify critically sampled images into point-like transients, streak-like transients and artifacts with high efficiency. However, we find these classification methods have different bias for classification of different targets. Based on the properties of these two methods, we propose to use the ensemble learning to further increase the classification accuracy. The classification accuracy of the ensemble method are more stable and higher than the CNN or the RNN method.\\
However, it should be mentioned that all the candidate transient images are obtained by the source extraction  method firstly (such as SExtractor) in this paper and also in other similar works. The source extract method has already placed a strong bias to the final classification results: the extracted sources are bigger than a predefined connected area and brighter than a predefined threshold. We find that some dim transients are missing during the source extract procedure and to obtain them, we need to modify the parameters of the SExtractor for every frame. To further increase the transient detection speed and accuracy, it is better to integrate the source extraction and candidate classification process and develop an object detection method, such as faster R-CNN \citep{Ren et al.(2015)}, YOLO \citep{Redmon et al.(2015)} or SSD \citep{Liu et al.(2015)}. Our future work will be developing the integrated object detection method with the ensemble neural network discussed in this paper.\\
\acknowledgments
The authors are grateful to the anonymous referee for his or her comments and suggestions, which have greatly improved the quality of this manuscript. The authors would like to thank the group from Xuyi Observation Station, Purple Mountain Observatory, for providing useful data for our work. The authors would like to thank Dr. Nan Li from University of Nottingham for providing very helpful suggestions for this paper. The authors would like to thank Dr. Alastair Basden from Durham University and Dr. Rongyu Sun from Purple Mountain Observatory for discussion of the data processing method. This work is supported by National Natural Science Foundation of China (NSFC)(11503018) and the Joint Research Fund in Astronomy (U1631133) under cooperative agreement between the NSFC and Chinese Academy of Sciences (CAS), Scientific and Technological Innovation Programs of Higher Education Institutions in Shanxi (2016033). Peng Jia is supported by the China Scholarship Council to study at the University of Durham. The code used in this paper is written in Python programming language (Python Software Foundation) with the package pytroch \url{https://github.com/pytorch}, astropy \citep{Astropy(2013)} and sklearn \citep{Buitinck(2013)}. 

\end{document}